\documentclass[aps,prl,twocolumn,showpacs,groupedaddress]{revtex4}
\usepackage{amsmath}
\usepackage[dvips]{graphicx}
\usepackage{amssymb}
\usepackage{color}
\newcommand{\T}{{\cal T}}

\newcommand{\V}{{\cal V}}

\begin{document}
\title{Tuning decoherence with a voltage probe}
\author{P. Roulleau}
\author{F. Portier}
\author{P. Roche}
\affiliation{Nanoelectronic group, Service de Physique de l'Etat Condens\'e,\\
CEA Saclay, F-91191 Gif-Sur-Yvette, France}
\author{A. Cavanna}
\author{G. Faini}
\author{U. Gennser}
\author{D. Mailly}
\affiliation{Phynano team CNRS, Laboratoire de Photonique et Nanostructures,\\
Route de Nozay, F-91460 Marcoussis, France}
\date{\today}
\begin{abstract}
We present an experiment where we tune the decoherence in a
quantum interferometer using one of the simplest object available
in the physic of quantum conductors : an ohmic contact. For that
purpose, we designed an electronic Mach-Zehnder interferometer
which has one of its two arms connected to an ohmic contact
through a quantum point contact. At low temperature, we observe
quantum interference patterns with a visibility up to 57\%.
Increasing the connection between one arm of the interferometer to
the floating ohmic contact, the voltage probe, reduces quantum
interferences as it probes the electron trajectory. This unique
experimental realization of a voltage probe works as a trivial
which-path detector whose efficiency can be simply tuned by a gate
voltage.
\end{abstract}
 \maketitle

Progress in nanofabrication techniques offers new opportunities to
study quantum effects in small size conductors. A remarkable
example has been the recent realization of electronic devices that
mimic the optical Mach-Zehnder interferometer. The properties of
these conductors have been successfully described using a "simple"
quantum scattering approach which considers electrons emitted by
"reservoirs" and scattered through the conductor. A limitation of
this so-called Landauer-B{\"u}ttiker theory is that it only treats
elastic scattering. Therefore, it cannot account for decoherence
or energy relaxation in electronic transport, a major issue for
real devices. This limitation has been cunningly circumvented by
theoreticians: they have introduced additional reservoirs whose
connection to the studied quantum circuit mimics the
decoherence\cite{Buttiker88IBMJRD32p63}. In these so-called
voltage probes, electrons loose their quantum phase memory by
thermalizing with the external world. Here we show the first
quantitative realization of a voltage probe with a small ohmic
contact which makes it possible to tune the decoherence in a
quantum interferometer.

A reservoir in the physics of quantum conductors is defined as
\textit{some} region of the conductor which absorbs all incoming
particles and emits "new" particles with a Fermi statistics at the
local electrochemical potential. Indeed, in the case of a sample
larger than the electronic coherence length, one cannot tell
exactly where are the reservoirs. They are simply assumed to be
located at the multiple extremities of the conductor under
consideration which exhibits quantum properties on a size scale
determined by the coherence length of excitations, or their energy
redistribution length. A voltage probe is a reservoir whose
precise position and coupling to the circuit determines the
location and the amount of decoherence.

The effect of a voltage probe can be explained in the following
manner: quasi-particles which have been probed by this additional
reservoir when going through the quantum conductor, loose their
phase so that nothing differentiates them from the electrons of
the reservoir. This theoretical construction is intimately linked
to which-path experiments, in the sense that when an electron is
absorbed by the additional reservoir, the ambiguity on the
particle's trajectory is lifted, suppressing interference effects.
Energy relaxation can also be described within the same framework
when the electrons are re-injected by the voltage probe into the
interferometer at thermal equilibrium. Indeed, in the case of the
electronic Mach-Zehnder interferometer, this approach has been
used to predict the current fluctuations in the presence of
decoherence or energy relaxation
\cite{Marquardt04PRL92n056805,Marquardt04PRB70n125305}.


We present here an experiment where a voltage probe introduces a
controlled energy redistribution. To this end, we have realized an
electronic Mach-Zehnder interferometer (MZI) operating in the
Quantum Hall Regime \cite{Ji03Nature422p415}. Here, the transport
occurs through one dimensional chiral channels located at the edge
of the sample (the edge states). These channels perfectly mimic
the photon beam and hence one can realize an electronic
counterpart to the optical interferometers. The voltage probe is
obtained with a small floating ohmic contact connected to one of
the arms of the interferometer through a controlled tunnelling
barrier (a quantum point contact (QPC)). Floating ohmic contacts
have already been used to enforce energy relaxation of noisy
currents \cite{Sprinzak00PRL84p5820,Oberholzer06PRL96n046804} but
without presenting an experimental set-up permitting the
exploration of their dephasing properties. More specifically, the
QPC allows us to tune the transmission probability $T_P$ towards
the voltage probe. As a result, the visibility of the quantum
interferences is reduced by a factor $\sqrt{1-T_P}$, which
represents the probability amplitude for a particle not to be
probed by the small floating ohmic contact.

\begin{figure}
\centerline{\includegraphics[angle=-90,width=9cm,keepaspectratio,clip]{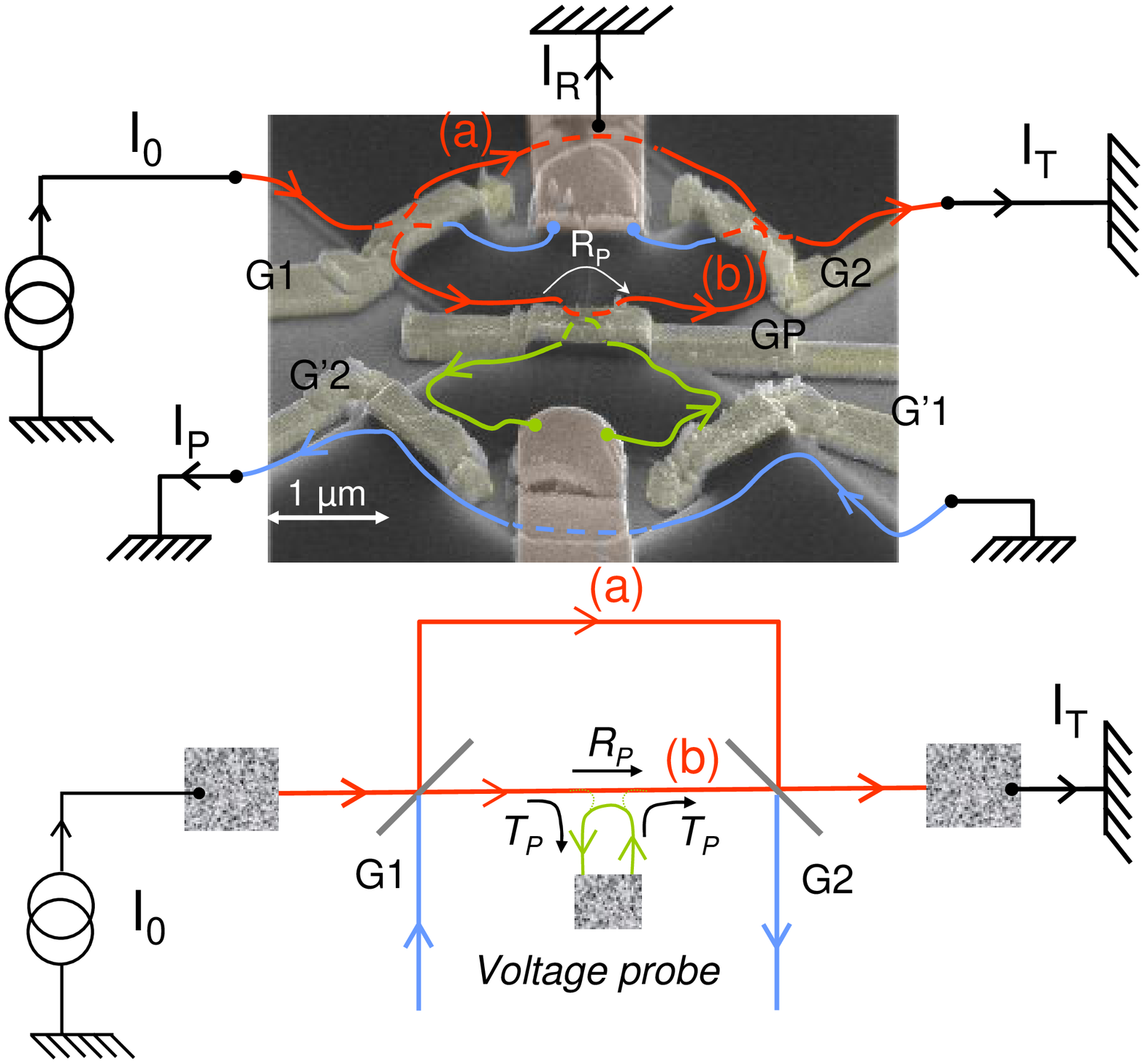}}
\caption{The experimental setup~: an electronic Mach-Zehnder
interferometer is designed by electron beam lithography on a high
mobility 2D electron gas in GaAs/GaAlAs heterostructure. One arm
(b) can be connected to a small floating ohmic contact which plays
the role of a voltage probe. QPCs G1 and G2 are the beam splitters
which split and recombine the particle trajectories. QPC GP allows
a control of the transmission probability $T_P$ toward the voltage
probe. G'1 and G'2 are additional QPCs which are either at pinch
off in the which-path experiment or, fully open to measure the
transmission through GP as a function of the gate voltage
$V_{GP}$. The top view is a colored tilted scanning electron
microscope view of the sample. The lines represent the edge
states. }\label{Principe.fig}
\end{figure}

A SEM view of our MZI is represented in figure (1). Starting from
a high mobility two dimensional electron gas in a GaAs/GaAlAs
heterostructure with a sheet density of
$n_S=2\times10^{11}$~cm$^{-2}$ and a mobility of $2.5\times
10^{6}$~cm$^2$/Vs, we patterned the geometry of the mesa, thus the
trajectory of the edge states, by e-beam lithography. The lengths
of arms (a) and (b) were both designed to be equal to 5.7~$\mu$m
yielding an enclosed area of 7.25~$\mu$m$^2$. In our MZI (see
figure (1)), there are 5 QPCs, G1, G2, G'1,G'2 and GP. G1 and G2
are the two beam splitters of the MZI itself, with transmissions
tuned to 1/2 to obtain a maximum visibility of the interferences
\cite{Roulleau07PRB76n161309}. GP, which is close to the
trajectory (b), has two purposes. In the pinch-off regime, it is
used to change the length of (b) in order to reveal the
interference pattern. In addition, GP serves as a connection
between (b) and the bottom small ohmic contact. We work at a
filling factor 2 at a magnetic field of 4.6~T giving rise to two
edge states. The inner one, not represented on figure
\ref{Principe.fig}, is fully reflected by G1 and G2.

We proceed as follows: we first fully open G'1 and G'2 to measure
the transmission trough GP ($T_P$) as a function of its voltage
$V_{GP}$. Once this reference obtained, we permanently close G'1
and G'2. The transmission probability through the MZI is measured
by a standard lock-in technique with an AC excitation $V_{AC}$=
1.2~$\mu$V smaller than $k_BT/e$, ensuring that the coherence
length of the source is only limited by the experimental
temperature of the order of 20~mK.

The interference pattern is revealed by varying either the
magnetic field or $V_{GP}$. Hence, GP both connects the trajectory
(b) to the voltage probe and sweep the phase difference between
the two arms of the MZI. In figure 2, a color plot of the
differential transmission versus the magnetic field and $V_{GP}$
is displayed. As one can notice, the amplitude of the oscillations
decreases as $V_{GP}$ increases, i.e. when the trajectories are
more connected to the voltage probe.

\begin{figure}
\centerline{\includegraphics[angle=-90,width=9cm,keepaspectratio,clip]{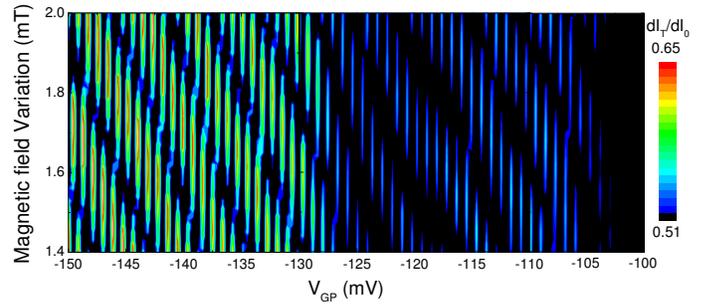}}
\caption{Color plot of the differential transmission $\T$ as a
function of the voltage probe gate voltage $V_{GP}$ and the
magnetic field. The color plot is set such that all the
transmissions lower than the mean transmission are in black. In
practice the visibility is measured by varying the magnetic field.
}\label{Pattern.fig}
\end{figure}

This visibility decrease is straightforward to understand. We call
$T_1$ and $T_2$ the transmissions through the beam splitters G1
and G2 and $T_P$ the transmission to the voltage probe. The
electron source injects an input current $I_0$ which has a
probability $I_T/I_0=T=t^*t$ to exit the MZI through the ohmic
contact located on the right side of figure 1. As we treat a
quantum circuit, $T$ is not the sum of the transmission
probability of the different trajectories $R_1R_2+T_1T_2$ (path
(a)+path(b)), but the squared sum of the transmission probability
amplitudes. The transmission amplitude $t$ through the MZI is then
the sum of three complex amplitudes corresponding to path (a),
path (b) and those multiple reflected paths (labelled by j) which
go through the small floating ohmic contact:
\begin{equation}
t=-r_1e^{i\phi_{a}}r_2+t_1r_Pe^{i\phi_{b}}t_2+t_1T_P\sum_j
(r_p)^je^{i\phi_{P_j}}t_2,
\end{equation}
$\phi_{P_j}$ being random phases accumulated in the voltage probe,
and $r_i$ and $t_i$ respectively stand for the reflection and
transmission coefficient of electronic wavefunctions by QPC $i$.
This leads to a transmission probability $T =
T_1T_2+R_1R_2-\sqrt{T_1R_2R_1T_2R_P}\cos[\phi_{a}-\phi_{b}]$,
where $R_i=|r_i|^2$, and $T_i=|t_i|^2=1-R_i$. The first two terms
of this expression correspond to the classical term whereas the
third one, which reveals the wave nature of electrons, oscillates
with the phase difference between the two arms. In the Quantum
Hall Regime, this is equal to the Aharonov-Bohm phase
corresponding to the magnetic flux threaded through the area
delimited by the two interfering trajectories. It can thus be
varied either by changing the enclosed area using GP or by
sweeping the magnetic flux \cite{Roulleau07PRB76n161309}. The
visibility of interferences defined as $\V
=(T_{MAX}-T_{MIN})/(T_{MAX}+T_{MIN})$ is:
\begin{equation}
\V=\V_0\times\sqrt{R_P}\label{vis}
\end{equation}
where $T_{MAX}$ and $T_{MIN}$ are the maximum and minimum
transmission respectively, $\V_0$ is the measured visibility
obtained when $T_P=0$. As expected, this means that only the part
of the wave function which does not go through the probe
contributes to the interferences. Equation \ref{vis} is thus a
consequence of the floating contact not affecting the mean
current: all the charges that have been absorbed into it are
re-injected into the circuit, so that the sum of the measured
transmitted current $I_T$ and of the current absorbed by the upper
small ohmic contact $I_R$ is conserved.

\begin{figure}
\includegraphics[angle=-90,width=8cm,keepaspectratio,clip]{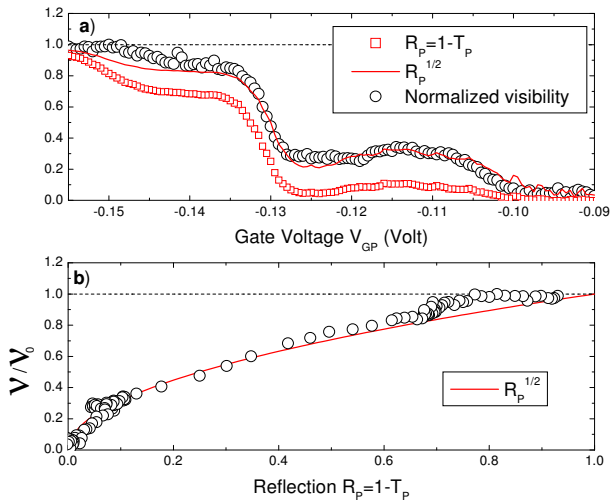}
\caption{\footnotesize{ Normalized visibility $\V/\V_0$ ($\V_0$ is
inferred for $T_P = 0$). \textbf{a)} $\V/\V_0$ (black circles),
$R_P$ (red squares) and $\sqrt{R_P}$ (red line) as a function of
$V_{GP}$. \textbf{b)} $\V/\V_0$ (black circles) as a function of
the measured $R_P$. The solid line is the $\sqrt{R_P}$ law
predicted by the theory.}}\label{buttiker.fig}
\end{figure}

In previous which path experiments using quantum conductors, the
dephasing occurred by coupling the electrons to a noisy
electromagnetic environment
\cite{Sprinzak00PRL84p5820,Buks98Nature391p871,Neder07PRL98p036803,Rohrlich07PRL98n096803}.
In our set-up, electrons re-emitted into the interferometer cannot
be distinguished from the other electrons of the probe. Reflecting
their interactions with the various degrees of freedom of the
floating contact, they bear a phase uncorrelated to the one of the
incident electrons. Hence, they do not contribute to the quantum
interferences that give rise to the Aharonov-Bohm term of the
transmitted current. To perform a quantitative analysis of the
voltage probe detection, we determined the transmission $T_P$ as a
function of $V_{GP}$. This is achieved by measuring $T_P =
dI_P/dI_0$ with $T_1=1$ and $T'_2=1$. The result is shown in
figure (3a). Then we closed G'1 and G'2 such that $I_P=0$. The
normalized visibility as a function of $R_P=1-T_P$ is plotted in
figure (3b). This is our main result, which shows the visibility
increasing as the square root of the reflection probability, in
perfect agreement with theory (Eq. (2)). It is noteworthy that
despite the small size of the ohmic contact (less than $1 \mu$
m$^2$), it shows no sign of Coulomb Blockade that would prevent
electrons from entering it and protect quantum interferences. This
is because the probe is connected through a metallic air bridge to
a much bigger bonding pad. This strongly increases its capacitance
and reduces its charging energy to a negligible level.

One can observe in figure (3a) that $R_P$ does not follow a
monotonous Fermi function like variation as predicted by the
saddle point model \cite{Buttiker90PRB41p7906}. There are two
resonances near $V_{GP}\sim -0.145$~V and $V_{GP}\sim -0.115$~V,
the first one ($R_P\sim0.75-0.9$) being associated with a
discrepancy between the observed visibility and the $\sqrt{R_p}$
law (see figure 3b). This is not the case for the second one. A
resonance whose trajectory is included in the MZI should in
principle be accompanied with a phase shift. As we will see, the
second resonance has such a phase shift but not the first one. It
means that around $V_{GP}\sim -0.145$~V, the measured conductance
is not directly related to $T_P$ when $G'1$ and $G'2$ are almost
closed.

Indeed, the phase variation $\delta\phi$ of the interferences
relates to the magnetic field variation $\delta B$ and $\delta
V_{GP}$ by $\phi=2\pi(\delta B.S+B\frac{dS}{dV_{GP}}.\delta
V_{GP})/\phi_0$, where $\phi_0$ is the quantum of flux $h/e$. This
phase variation leads to tilted black regions in figure 2, given
by $\delta B.S \propto \frac{dS}{dV_{GP}}.\delta V_{GP}$. At the
resonance which appears in the measurement of $R_P$ for
$V_{GP}\sim -0.115$~V, the separation between the tilted region is
no longer regular, indicating that when crossing the resonance a
additional phase shift appears in the interferences
\cite{Avinun05Nature436p529}. Inspecting in detail the conductance
trace for a given magnetic field as a function of $V_{GP}$ we
found a phase shift of approximately $\sim\pi$ around this
resonance, although our phase measurement is not precise enough to
determine the exact shape of the phase variation. The absence of
such phase shift in the other resonance close to $V_{GP}\sim
-0.145$~V explains the small discrepancy with the $\sqrt{R_P}$ law
observed: when measuring $R_P$, all the closed trajectories at a
distance closer than the coherence length from GP
\cite{Roulleau08PRL100n126802,Roulleau08PRL101n186803} could
possibly lead to resonances. Here, we are in the case where the
closed trajectory leading to this resonance is outside the MZI
when G'1 and G'2 are at pinch off. Hence the value of the measured
$R_P$ is not what should be taken into account for the visibility
decrease.

To summarize, we have shown that a small floating ohmic contact is
a voltage probe that can be used to destroy quantum interferences
in a controlled way. For that purpose, we have used a QPC to drive
the amplitude probability of the absorbtion of an electron in the
voltage probe. Then, via interference measurements, we have proved
that electrons absorbed and re-emitted by the probe acquire a
random phase and do not contribute to the interference process.
This work opens new possibilities regarding the study of the
voltage and dephasing probe, the most promising being its full
counting statistics, as recently proposed
\cite{Pilgram06PRL97n066801,Forster07PRB75n035340,Forster07NJP9p117}.


\end{document}